\documentclass[traditabstract]{aa}
\usepackage{graphicx}
\usepackage{rotating,subfigure,amssymb,afterpage}
\usepackage{txfonts}
\usepackage{natbib}
\usepackage{color}
\usepackage[pdftitle={Cluster Magnetic Fields},
pdfauthor={H. B\"ohringer et al.},
pdfsubject={}, pdfkeywords={}, bookmarks, bookmarksopen, 
colorlinks=true, linkcolor=blue, citecolor=blue, anchorcolor=blue,
urlcolor=blue]{hyperref} 
\newcommand{\propsim}{\lower 3pt \hbox{$\, \buildrel {\textstyle
      \propto}\over {\textstyle \sim}\,$}}
\begin{document}

   \title{The Cosmic Large-Scale Structure in X-rays (CLASSIX) \\
   Cluster Survey I: Probing galaxy cluster magnetic fields\\ 
   with line of sight rotation measures}

   \author{Hans B\"ohringer\inst{1}, Gayoung Chon\inst{1} and Philipp P. Kronberg\inst{2}}

   \offprints{H. B\"ohringer, hxb@mpe.mpg.de}

   \institute{$^1$ Max-Planck-Institut f\"ur extraterrestrische Physik,
                   D-85748 Garching, Germany.\\
              $^2$ Department of Physics, University of Toronto, 60 St George Street, 
                   Toronto, ON M5S 1A7, Canada
}

   \date{Submitted 08/05/16}

\abstract{To search for a signature of an intracluster magnetic field,
we compare measurements of Faraday rotation of polarised extragalactic 
radio sources in the line of sight of galaxy clusters with those outside.
To this end, we correlated a catalogue of 1383 rotation measures of extragalactic 
polarised radio sources with galaxy clusters from the 
{\sf CLASSIX} survey (combining {\sf REFLEX II} and {\sf NORAS II}) 
detected by their X-ray emission in 
the ROSAT All-Sky Survey. The survey
covers 8.25 ster of the sky at $|b_{II}| \ge 20^o$. We compared the rotation 
measures in the line of sight of clusters within their projected radii of $r_{500}$
with those outside and found a significant excess of the dispersion of the
rotation measures in the cluster regions. Since the observed rotation measure
is the result of Faraday rotation in several presumably uncorrelated
magnetised cells of the intracluster medium, the observations correspond
to quantities averaged over several magnetic field directions and strengths.
Therefore the interesting quantity is the dispersion or standard deviation
of the rotation measure for an ensemble of clusters. In the analysis
of the observations we found a standard deviation of the rotation measure
inside $r_{500}$ of about 120 ($\pm 21$) rad m$^{-2}$. This compares to about 
56 ($\pm 8$) rad m$^{-2}$ outside. 
Correcting for the effect of the Galaxy with the mean rotation measure 
in a region of 10 deg radius in the outskirts
of the clusters does not change the outcome quoted above. 
We show that the most X-ray luminous 
and thus most massive clusters contribute most to the observed excess 
rotation measure.
Modelling the electron density distribution in the intracluster medium 
with a self-similar model based on the REXCESS Survey, 
we found that the dispersion
of the rotation measure increases with the column density, and we deduce a 
magnetic field value of about  $ 2 - 6~ (l/10kpc)^{-1/2} \mu$G assuming a 
constant magnetic field strength, where
$l$ is the size of the coherently magnetised intracluster medium cells. This 
magnetic field energy density amounts to a few percent of the average 
thermal energy density in clusters. On the other hand, when we
allowed the 
magnetic field to vary such that the magnetic energy density is a constant
fraction of the thermal energy density, we deduced a slightly lower value 
for this fraction of $3 - 10~ (l/10kpc)^{-1/2}$ per mille. Compared to the
situation in the Milky Way, where the ratio of the magnetic to
thermal energy density is about unity, this ratio is much lower
in galaxy clusters. The reason for this is most probably the different
generation mechanism for the magnetic field, which is mostly powered
by supernovae in the Galaxy and by turbulence from cluster mergers
in galaxy clusters. The latter process 
sets a natural upper limit on the growth of the
magnetic field.
} 


\authorrunning{B\"ohringer et al.}
\titlerunning{Galaxy cluster magnetic fields from rotation measures}
   \maketitle
%

\section{Introduction}

In the past decades we have gained much insight into the structure and
physics of galaxy clusters from observations of their hot intracluster
medium (ICM), which enabled us among other things to measure their masses, 
characterise their dynamical state, and use them to study the large-scale 
structure of the Universe (e.g. Voit 2005, B\"ohringer \& Werner 2010, 
Kravtsov \& Borgani 2012, Chon et al. 2012). In most of these studies 
the ICM was treated as a thermal plasma, and magnetic fields were ignored. 
On the other hand, we know 
that magnetic fields are present, mostly through the observation
of synchrotron radiation and Faraday rotation of polarised radio signals
from background radio sources. The radio synchrotron emission comes
from cosmic-ray electrons in the magnetic field of the ICM
observed in radio halos and radio relics in clusters (e.g. Kim, et al. 1990,
Feretti et al. 2012). The Faraday rotation measure (RM) is imprinted on the signal of
polarised radio sources seen in the background in the line of sight 
of galaxy clusters (e.g. Kim, Tribble \& Kronberg 1991, 
Feretti et al. 1995, Clarke et al. 2001). Thus
a magnetic field seems to be a ubiquitous component of the ICM
(e.g. Carilli \& Taylor 2002),
and for many physical processes its presence needs to be taken
into account. In this paper we use the statistics of RMs
in the lines of sight through galaxy clusters to infer properties of the 
cluster magnetic fields.

Polarised electromagnetic radiation traversing a magnetised
plasma is subject to Faraday rotation, with a RM
given by

\begin{equation}
RM ~=~ 811.9~ \left({ n_e \over 1 {\rm cm^{-3}}} \right)~ 
  \left( {B_{||} \over 1  {\rm \mu G}} \right)~
  \left( {L \over 1 {\rm kpc}} \right)~ {\rm rad~m}^{-2}~~~~.
\end{equation}

For an ICM with typical electron densities
of $10^{-3}$ cm$^{-3}$, a size of about one Mpc, magnetic fields of a few
$\mu$G, and a magnetic field aligned with the line of sight,
we would expect RMs with values of about 1000 rad~m$^{-2}$.
Since the magnetic field is most probably tangled, the
effect partly averages out, and for an ordering scale of the 
magnetic field of about 10 kpc, we expect an
RM dispersion at least
an order of magnitude smaller than this value. Owing to the
frequency dependence of the effect, the rotation of the 
polarisation angle can be determined
when the polarisation at more than one frequency is measured.
For expected RM values of about 100 rad~m$^{-2}$ , the
rotation of the polarisation vector at a
wavelength of 21 cm, for instance, is 4.42 rad. Measurements at
several frequencies are therefore required to derive
the RM unambiguously. Observed RM signals 
in the line of sight of clusters also show, in addition to the effect
of the ICM, the imprint of the galactic magnetic field
and interstellar medium and possibly a source-intrinsic
RM. Typical values for the galactic RM
are about 50 rad~m$^{-2}$
(Simard-Normandin \& Kronberg 1980, Simard-Normandin, Kronberg
\& Button 1981). This means that there is a high possibility to observe the
imprint of the ICM RM with a statistical sample
of RMs in galaxy cluster sight-lines. 

The first attempt of such a study was conducted by 
Lawler \& Dennison (1982). They found a signature of excess 
RMs in clusters with about 80\% confidence for 12
sight-lines with impact parameters of $ < 1/3$ Abell radius
and 12 sight-lines
at $1/3 - 1$ Abell radii. In their analysis the authors concluded
that the typical magnetic field strength is about $0.07~ N^{1/2}
~ \mu$G with an upper limit of 0.2 $\mu$G, where $N$ is the number
of ICM cells with a coherent magnetic field.
Kim, Tribble \& Kronberg (1991), Feretti et al. (1995), 
and subsequently
others studied the magnetic field in the Coma cluster
with RMs along several sight-lines and partly
also with RMs of extended radio sources, concluding
a magnetic field strength of $ <\sim 7\mu$G and 
an ordering scale of 1 - 10 kpc. Clarke et al. (2001) studied
the statistical effect of cluster ICM for 27 sight-lines through
X-ray luminous galaxy clusters, and found a clear signal.
The galaxy clusters in their sample all had good enough X-ray 
observations to allow an individual modelling of the properties
of the cluster ICM to determine the electron column density
in the line of sight. From comparison of the RMs
detected in the cluster sight-lines with those measured outside
and an analysis of the electron column densities, they
deduced an average magnetic field strength of  
$<|B|> = 5 - 10~ (l/10 {\rm kpc})^{1/2}~ h^{1/2}_{75} \mu$G. 
Since then, several detailed studies of RMs
for multiple sight-lines through individual clusters have also been 
conducted (e.g. Feretti et al. 1999, Murgia et al. 2004,
Feretti et la. 2012). A search for magnetic fields on
even larger scales has been conducted by Xu et al. (2006),
who found a signal of enhanced RM in the Hercules and
Perseus-Pisces supercluster implying a magnetic field 
of $\sim 10^{-7}$ G with an upper limit of $\sim 3 \times 10^{-7}$ G.
Kronberg et al. (2007) also found faint radio emission on scales of 
several degrees (up to 4 Mpc) close to the Coma cluster in the Great Wall region,
implying magnetic fields of about $2 - 4 \times 10^{-7}$ G.  

Meanwhile, the number of extragalactic RMs has 
increased since the study of Clarke et al. (2001), 
and our catalogue of X-ray detected galaxy clusters
with well-understood properties is also much larger, therefore
we now
revisit the statistical detection of ICM 
RMs. With the better statistics we can study
the correlation of the observed RM with cluster properties in more detail 
and produce more quantitative results.
In total, we now have RMs from 92 sight-lines through
clusters, which increases the statistics by about a factor of four.     

To determine all distance-dependent parameters, we use
a flat $\Lambda$CDM cosmology with $H_0 = 70$ km s$^{-1}$
Mpc$^{-1}$ and $\Omega_m = 0.3$. All X-ray luminosities are quoted in the
$0.1 - 2.4$ keV band.

\section{Observational data}

\subsection{ X-ray galaxy cluster sample}

The galaxy cluster sample {\sf CLASSIX} that we used here combines 
for the first time the northern 
and southern clusters that are identified by their X-ray emission in
the ROSAT All-Sky X-ray Survey (Tr\"umper 1993, Voges et al. 1999).
The southern sample, {\sf REFLEX II} (B\"ohringer et al. 2004, 2013, 
Chon \& B\"ohringer 2012), covers  
the southern sky below declination +2.5$^o$ and at Galactic
latitude $|b_{II}| \ge 20^o$. 
The survey region, source detection, galaxy cluster sample definition and compilation,
construction of the survey selection function, and tests of the completeness of the
survey are described in B\"ohringer et al. (2013). Known galaxy clusters
in the region of the Magellanic Clouds, excised in {\sf REFLEX II},
are included here. The northern sample, 
{\sf NORAS II} (B\"ohringer et al., 2000, B\"ohringer et al., in preparation),
has been constructed in the same way as {\sf REFLEX II}. 
The two samples can thus be
combined (taking into account the overlap region at declination
0 to +2.5$^o$) and described with a common selection function.
In total, the catalogue contains 1722 X-ray luminous galaxy
clusters.

In summary, the overall survey area is $ \sim 8.25$ ster. The nominal flux-limit 
down to which galaxy clusters have been identified in the RASS in 
this region is $1.8 \times 10^{-12}$ erg s$^{-1}$ cm$^{-2}$ in the
0.1 - 2.4 keV energy band. The nominal flux limit imposed on the survey 
was calculated from the detected photon count 
rate for a cluster X-ray spectrum characterised by a temperature of 5 keV, 
a metallicity of 0.3 solar,
a redshift of zero, and an interstellar absorption column
density derived from the 21cm sky survey described
by Dickey and Lockman (1990). This count rate is analogous 
to an observed object magnitude corrected for Galactic extinction 
in the optical.

Spectroscopic redshifts have been obtained for all the clusters,
except for 25 missing redshifts in {\sf NORAS II}, which are excluded here.
Based on these redshifts, proper fluxes and X-ray luminosities
were calculated iteratively, by using the proper K-correction
for the given redshift and assuming an intracluster temperature
according to our X-ray luminosity - temperature relation.
This relation was determined with the {\sf REXCESS} cluster survey 
(B\"ohringer et al. 2007, Pratt et al. 2009), a representative 
subsample of the survey clusters. 

The most important cluster characterisations are
the cluster mass and the 
fiducial cluster radius. We use $r_{500}$ here for
the cluster outer radius, which is defined as the radius inside 
which the mean mass density of a cluster
is 500 times the critical density of the Universe.
$r_{500}$ and $M_{500}$ (the mass inside $r_{500}$) are determined from 
the relation of the X-ray luminosity and cluster mass as described
in B\"ohringer et al. (2013), with the following luminosity mass
relation:

\begin{equation}
M_{500} ~=~ 2.48~ L_{X,500}^{0.62}~ E(z)^{-1}~~~~,
\end{equation} 

\noindent
where $E(z)^2 = \Omega_m (1+z)^3 ~\Omega_{\Lambda}$.

\medskip
X-ray emission is observed in most clusters
in the ROSAT Survey and in follow-up XMM Newton observations
up to a radius close to $r_{500}$. Thus the characterisation of the cluster
ICM is quite reliable out to this radius. 
The mean typical X-ray luminosity of the cluster sample is
$2.3 \times 10^{44}$ erg s$^{-1}$ and the mean cluster mass is about 
$3 \times 10^{14}$ M$_{\odot}$.

\subsection{Sample of rotation measures}

To probe the Faraday RMs imprinted by clusters on the line of
sight
to polarised radio sources, we have  drawn from the sample of Faraday rotation
measures shown in Fig. 1 of Kronberg \& Newton-McGee (2011). 
Selecting only measurements at locations
with $|b_{II}| \ge 20^o$ and removing all known redshifts 
below $z = 0.05$, which is intended to exclude Galactic sources, leaves 1383 RMs 
in the sample. Radio sources with double lobes with separate
RM determinations are treated as separate sources.
Because of the more complex Galactic foreground structure we decided not
to extend our survey into the Galactic band $|b_{II}| < 20^o$.

We also considered using the RM dataset of Taylor et al. (2009).
Apart from the fact that the RM were obtained from only two frequencies,
which can lead to ambiguities for high values of RM, we only found two
RM sightlines overlapping with clusters for this sample.   

Radio wavelengths used in the RM determinations ranged from 
$\lambda \sim  2$ cm to $\sim 31$ cm, but this varied from source to source
depending on the radio telescopes and available wavelengths. 
They are an expanded and updated 
version of the 555 RM sample of Simard-Normandin et al. (1981) 
and were derived from polarisation measurements at many wavelengths, 
using methods described in detail by Simard-Normandin et al. (1981). 
The large available baseline in $\lambda^2$ gives an unprecedented 
average precision in the RM determinations
(usually $< \pm 2$ rad m$^{-2}$) (see Pshirkov et al. 2011).

\section{Data analysis}

To search for an imprint of RM that is due to clusters of galaxies,
we sorted the polarised radio sources with RMs by
their distance to the nearest cluster in the sky. As a fiducial
radius for the edge of the galaxy clusters we took $r_{500}$.
The value of $r_{500}$ was calculated using the cluster mass 
obtained  from the empirical relation of X-ray luminosity 
and cluster mass given above, following B\"ohringer et al. (2013):

\begin{equation}
r_{500} ~=~ 0.957~ L_{X,500}^{0.207}~ E(z)^{-1}~~~~~.
\end{equation}

Outside of $r_{500}$ in the outskirts of clusters, the column 
density of the ICM is too low to 
expect a significant excess RM. The integrated cross section
of all clusters in the sample in terms of sky area amounts 
to 203.8 deg$^2$ (0.062 ster, 0.75\% of the entire survey area). 
In total, we found 92 
radio sources with known RMs inside the projected cluster
locations. Comparison of the radio source and cluster redshifts shows 
that most of the polarised
radio sources are radio galaxies in the clusters. Only 10 are
background sources with known redshifts, and for 26 polarised radio sources
the redshifts are unknown. We assume that for the radio galaxies in
clusters the ray path intersects on average about half of the 
ICM column density. Thus we
can include them as probes for the ICM magnetic field. 

Figure 1 shows the 
observed RMs as a function of cluster-centric distance
scaled to $r_{500}$. We note an obvious enhancement of the scatter
of RMs inside $r_{500}$ compared to polarised radio sources at larger
distances from the cluster centres. The plot also shows that even though
most of these radio sources are hosted in the cluster, the 
majority are not the central dominant cluster galaxies, which 
might be located in cooling cores with enhanced magnetic fields,
producing particularly high RMs.
Calculating the standard deviation of the 
RM (in the following referred to as the scatter) 
inside $0.5 r_{500}$,
$r_{500}$, and at $ 0.5 r_{500} < r < r_{500}$
and $ r_{500} < r < 10 r_{500}$, we found the following 
values of the scatter for the RMs: 
 $123 \pm 21$ ,$120 \pm 21$, $144 \pm 43$, and $57 \pm 6$ rad m$^{-2}$,
respectively. These striking results are also summarised in Table 1.
The uncertainty of the scatter was determined from
1000 bootstrap simulations.
One particularly high RM inside $ r_{500}$
has a value of 657.3 rad m$^{-2}$. 
It is located inside a very massive cluster, Abell 1763,
with an X-ray luminosity of $9.2 \times 10^{44}$ erg s$^{-1}$
corresponding to a mass of $8.3 \times 10^{14}$ M$_{\odot}$. 
The radio source itself is a double-lobed wide-angle tail 
radio galaxy 4C +41.26 in the central region of the cluster
(Owen \& Ledlow 1997).
While there is a chance that the high RM originates locally 
from the radio source, the fact that it is correlated with a
rare high-mass cluster suggests a high probability that the 
RM is associated with the cluster. 
Removing this source from the calculations by imposing an RM cut
$|RM| < 500 $ rad m$^{-2}$, we obtain the following values
for the scatter: $94 \pm 13$, $100 \pm 15$, $114 \pm 43$, and 
$57 \pm 6$ rad m$^{-2}$, respectively. The signal for the 
clusters has decreased, but is still significantly stronger inside
the clusters compared to outside.

Figure 2 also shows the distribution of the RMs
as a function of physical radius in units of Mpc for a better
comparison with our earlier study reported in Clarke et al (2001). Using
physical radii, the signal is also very clear, and the scatter
is $115 \pm 27$, $111 \pm 21$, $68 \pm10$  rad m$^{-2}$ 
for impact parameters $< 0.5$ Mpc, $< 1$ Mpc, and outside 1 Mpc, 
respectively.

\begin{figure}
   \includegraphics[width=\columnwidth]{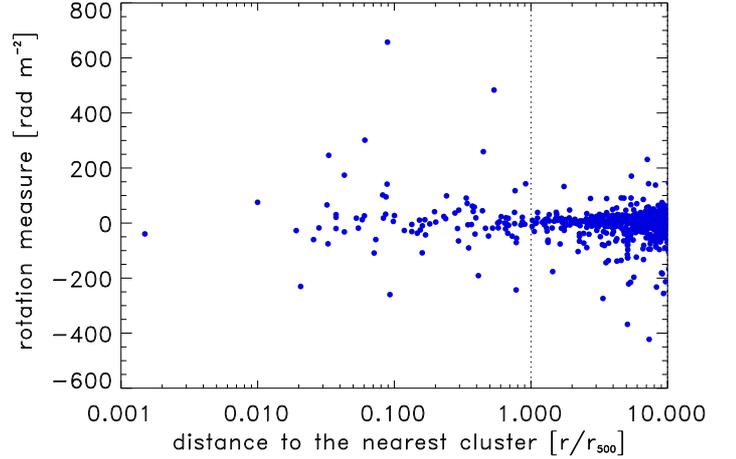}
\caption{Rotation measure as a function of cluster-centric distance
scaled to the fiducial cluster radius, $r_{500}$.
The vertical dashed line shows a cluster radius of
$r_{500}$. 
}\label{fig1}
\end{figure}

\begin{figure}
   \includegraphics[width=\columnwidth]{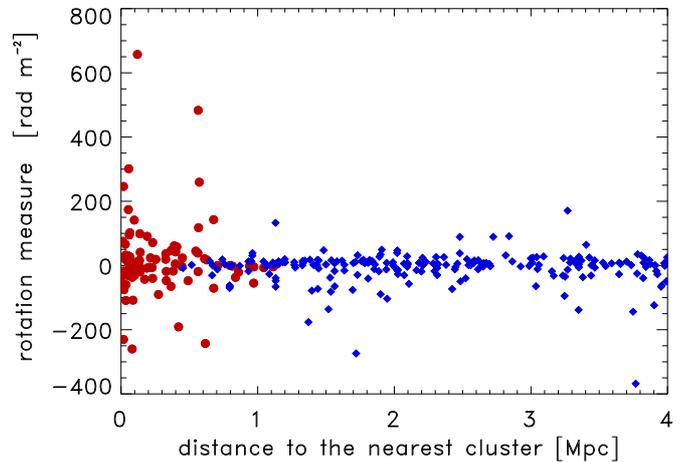}
\caption{Rotation measures
as a function of cluster-centric distance in physical units (Mpc). 
The rotation measures
inside $r_{500}$ are marked by red circles and those outside by blue diamonds.
}\label{fig2}
\end{figure}

   \begin{table}
      \caption{Standard deviation of the observed rotation measure in
      different regions in and around the line of sight of galaxy clusters.
     The different analyses are explained in the notes below. $N_e$ refers
     to the electron column density in the line of sight.}
         \label{Tempx}
      \[
         \begin{array}{lrrrr}
            \hline
            \noalign{\smallskip}
{\rm } & ~< 0.5 r_{500}  & ~~~0.5 - 1 r_{500} &~~< r_{500} 
   & ~~1 - 10 r_{500} \\
            \noalign{\smallskip}
            \hline
            \noalign{\smallskip}
{\bf A} & 123 \pm 21 & 114 \pm 43 & 120 \pm 21 & 57 \pm 6 \\
{\bf B} &  94 \pm 13 & 114 \pm 43 & 100 \pm 15 & 57 \pm 6 \\
{\bf C} & 124 \pm 21 & 112 \pm 43 & 120 \pm 21 & 52 \pm 6 \\
            \noalign{\smallskip}
            \hline
            \noalign{\smallskip}
{\rm } & < 0.5 {\rm Mpc} & 0.5 - 1  {\rm Mpc}  & < 1 {\rm Mpc} & >1{\rm Mpc}  \\
            \noalign{\smallskip}
            \hline
            \noalign{\smallskip}
{\bf D} & 115 \pm 27 & 107  \pm 33 & 111 \pm 21 & 68 \pm 10 \\
            \noalign{\smallskip}
            \hline
            \noalign{\smallskip}
N_e & < 6.5 \cdot 10^{20}  & ~~6.5~- &15.2 \cdot 10^{20}~~~~~~ & > 6.5 \cdot 10^{20} \\
            \noalign{\smallskip}
            \hline
            \noalign{\smallskip}
{\bf E} & 58.6 \pm 10.5 &  122.2 & \pm 35.6~~~~~~~~~~~~ & 157.0 \pm 40.3 \\
            \noalign{\smallskip}
            \hline
            \noalign{\smallskip}
         \end{array}
      \]
{\bf Notes:} {\bf A} RMs uncorrected for the galactic contribution,~ 
{\bf B} same as A 
 without RMs $| RM | > 500$ ,~  
{\bf C} with RMs corrected for Galactic contribution,~  
{\bf D} region defined in physical radii,~
 {\bf E} RMs as a function of electron column density given
 in units of cm$^{-2}$. The caption for the last line gives the range of electron
column densities for the bin. The bin boundaries have been chosen such 
that each bin has a similar number of RMs. 
\label{tab1}
   \end{table}
%

The clusters in our sample cover a wide mass range from 0.02 to $19.1
\times 10^{14}$ M$_{\odot}$. Smaller clusters will have a lower electron
column density of the ICM in the line of sight for a given radius scaled
by $r_{500}$. This indicates that most of the signal comes from the 
most massive clusters. To test this, we separated
the cluster sample into two halves split by the median X-ray luminosity
of $0.41 \times 10^{44}$ erg s$^{-1}$ , which corresponds to a mass of
about $1.4 \times 10^{14}$ M$_{\odot}$.
Above this median $L_{X}$ , the scatter of the RM is $158 \pm 34$ rad m$^{-2}$
, while for the other half of the sample with lower X-ray luminosity the
scatter in the RM is $62 \pm 11$ rad m$^{-2}$. Thus we see a clear
sign of the effect, confirming that the observed excess scatter
in the RM in the lines of sight of galaxy clusters 
is due to the cluster ICM. We also checked the redshift
distribution of the clusters contributing to the observed excess RM.
We find that while the median redshift of the ROSAT cluster sample
is about $z = 0.1$, only nine clusters above this redshift have an observed
RM. This is probably due to the decreasing 
apparent size of the clusters
with increasing redshift. The 92 RMs found are projected
onto 65 clusters. In most clusters only one sight-line with known 
RM is found. A notable exception is the nearby Coma cluster at 
a distance of $\sim 100$ Mpc, where 12 RMs are in our catalogue
and more are known (Kronberg 2016).

\section{Results and discussion}

Part of the observed RM in the line of sight of clusters comes from the 
effect of the foreground interstellar medium in our Galaxy. We aimed to
correct for this foreground effect by removing the average RM
signal in the surroundings of the clusters. To demonstrate the usefulness
of such a correction, we studied the distribution of the RMs
in the outskirts of clusters at radii $r_{500}~ <~ r~ <~ 10$ deg. 
In this analysis RMs in the lines of sight to clusters were
excluded.

Figure 3 shows the distribution of the RMs in sky regions
surrounding the clusters as a function of the mean RM in each
region containing these sight-lines. On average, 21 RMs
are found in each of the outskirt regions. The figure shows 
that the RM scatter is quite large. Tt is in general
larger than the mean value we wish to use to correct for
the Galactic RM contribution.
However, the individual
RMs are also clearly correlated with the mean. This clear correlation implies that 
we can improve on the measurement of the extragalactic RMs
by subtracting the mean of the foreground RM that is detected
in regions excluding clusters. To be conservative, we included
the scatter in the determination of the average RM in the
uncertainty of the corrected extragalactic RM. 

This suggests a correction
for the foreground emission measure by subtracting the mean
of the foreground signal seen in the cluster outskirts from the 
observed cluster RM. The large RM scatter
in each background area is then accounted for by including
the scatter of the uncertainty of the corrected cluster RM. 

\begin{figure}
   \includegraphics[width=\columnwidth]{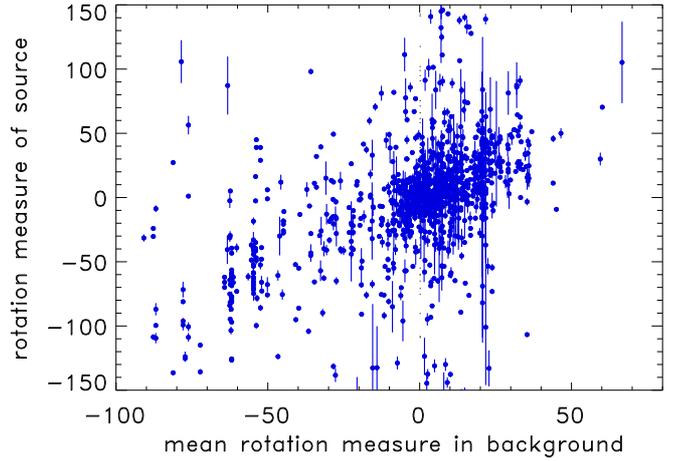}
\caption{Rotation measures in the outskirts of all clusters in our sample
at a radius $> r_{500}$ and within 10 Mpc as a function of the mean rotation
measure in each of the regions. Lines of sight falling into one of
the clusters in our sample have been excised.  
}\label{fig3}
\end{figure}

Figure 4 shows the RMs corrected in this way as a function of the cluster-centric 
radius scaled to $r_{500}$. The signal barely changes in comparison to 
the uncorrected data shown in Fig. 1, as can also be seen in Table 1.
We have also studied how the correction for Galactic contributions changes
when we decrease the size of the background region around the cluster.
Changing the outer
radius of the background region between 5 and 10 deg causes differences in
the results that are much smaller than the uncertainties. 

\begin{figure}
   \includegraphics[width=\columnwidth]{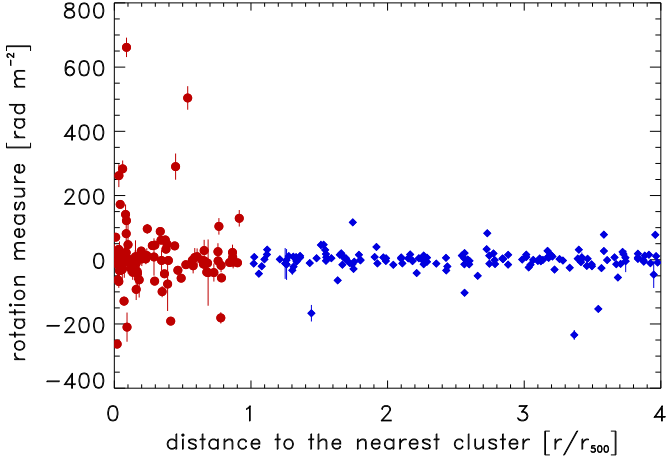}
\caption{Corrected rotation measures as a function of cluster-centric radius
scaled by $r_{500}$.
}\label{fig4}
\end{figure}

\begin{figure}
   \includegraphics[width=\columnwidth]{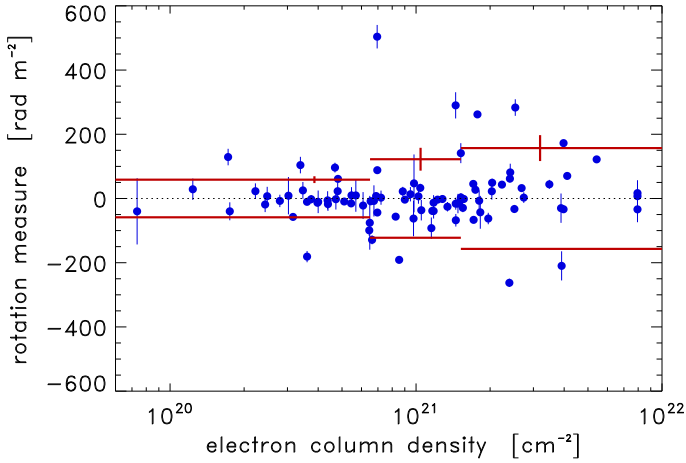}
\caption{Rotation measure as a function of electron column density
of the ICM in the sight-line. The red bars give the dispersion
of the rotation measures in three bins with error bars
shown only on the positive side. 
}\label{fig5}
\end{figure}

To obtain more quantitative results as a function of the physical properties 
of the cluster ICM, we determined the electron column density of the ICM
in the line of sight of the radio sources and inspected the 
RMs as a function of the electron column density according to
Eq. 1. To model the ICM density distribution of the clusters, we
assumed that the clusters are spherically symmetric and can be described
by a self-similar model, scaling with the cluster mass or X-ray 
luminosity. We based the model on the results of our study of the
REXCESS sample (B\"ohringer et al. 2007, Croston et al. 2008). 
We used the following
parameterised function, which describes the electron
density profile well:

\begin{equation}
n_e(r)~ =~A~ \left( r \over r_c \right)^{ \alpha}~ 
\left[ 1 + \left( r \over r_c \right)^2~
\right]^{-{3\beta \over 2} + {\alpha \over 2}}~~ . 
\end{equation}

This functional form was now fitted to the electron density profiles of the
{\sf REXCESS} sample clusters by scaling the radii to $r_{500}$
and applying self-similar scaling. The resulting best-fitting parameters are $\alpha = 0.41$ 
and $\beta = 0.64$. The normalisation of the function is consistent
with a gas mass fraction of the clusters of about 10\%.
For each line of sight we integrated the electron density out to
$r_{500}$. Since several radio sources sit inside
clusters, we integrated the column densities for these systems
only over the half sphere, whereas we used the full sphere for
the rest, including the radio sources without known redshifts.  

Figure 5 shows the results of the corrected RMs as a function
of estimated electron column density. In addition, the plot shows 
the RM scatter in three bins of the column density,
with values of $58.6 \pm 10.5$, $122.3 \pm 35.6$, and $157.0 \pm 40.3$
rad m$^{-2}$.
Again the RMs clearly increase with electron column
density in the line of sight. The uncertainties for the scatter were
obtained with 1000 bootstrap simulations in each case.
The scatter in the lowest bin of the electron column density
is almost identical to that in the surroundings of the clusters.
This is not surprising given the low electron column densities in this
bin, which can be taken as a baseline for the effect of the Galactic 
foreground. We subtracted it from the observed values
in the other two bins in quadrature.

These results permit us to estimate the magnetic field by means
of Eq. 1. Defining the electron column density
$N_e = n_e \times L$ and assuming in a first step that the 
magnetic field is ordered on cluster scale, we obtain
\begin{equation}
\left({ B_{||} \over 1 \mu {\rm G}}\right)~ = 3.801 \times 10^{18}~ 
\left( {RM\over {\rm rad~m}^{-2}} \right)~  
\left( {N_e\over {\rm cm}^{-2}}\right)^{-1} ~~~~.  
\end{equation} 

Now we have to consider that the observed RM originates in
the superposition of 
many ICM plasma cells in the line of sight with different magnetic
field orientations. The RM will thus be diluted
by averaging over all cells in the line of sight by a factor  
of $\Lambda = (L/l)^{1/2}$, where $L$ is the length of the ICM 
column and $l$ is the typical size of the plasma cells with coherent 
magnetic field direction. We can then calculate the 
line of sight magnetic field strength as

\begin{equation}
\left({ B_{||} \over 1 \mu {\rm G}}\right)~ = 3.801 \times 10^{18}~
\left( {\sigma (RM)\over {\rm rad~m}^{-2}} \right)~  
\left( {N_e\over {\rm cm}^{-2}}\right)^{-1}~\Lambda  ~~~~~~~,  
\end{equation}   

\noindent
where $N_e$ in the electron column density in the line of sight. With
values for the mean $N_e$ in the second and third bin in Fig. 5
($1.04 \times 10^{21}$ and $3.19 \times 10^{21}$ cm$^{-2}$)
and assuming typical values of $L \sim 1$ Mpc and $l \sim 10$ kpc,
we find for the line of sight magnetic field component values
of $0.38~ (\pm 0.13) \times \Lambda$ and 
$0.17~ (\pm 0.04)\times \Lambda  ~\mu$G. Combining the
two bins yields a value of $0.18~ (\pm 0.05)\times \Lambda ~\mu$G.
Typical coherence lengths of the magnetic field have been found
to be in the range 2 - 25 kpc
(e.g. Feretti et al. 1999, Govoni et al., 2001, 
Taylor et al. 2001, Eilek \& Owen 2002, Murgia et al. 2004).
This is interestingly similar to the value measured by Kim et al.
(1990) from the projected RM variation along the cluster-internal
extended radio source 5C4.81.  
Consequently, we scaled our results to $l = 10$ kpc. 
Assuming that the magnetic field is isotropic globally,
an average column length of 1 Mpc, and a cell size of about 10 kpc,
we find a total magnetic field a value of $<|B|> \sim 3 ({+3 \atop -1}) 
\times (l/10kpc)^{-1/2} ~\mu$G. This value confirms our previous results reported in 
Clarke et al. (2001) and agrees well in general with values quoted in the literature for the
magnetic fields on global scales from equipartition considerations
of radio halos and Faraday rotation measurements (e.g. Feretti et al. 2012). 

We can further refine our model by allowing the magnetic field to
vary within the cluster. A reasonable assumption is that the magnetic
energy density has a constant ratio to the thermal
energy density (e.g. Miniati 2015 and Miniati \& Beresnyak 2015).
This results from a dynamo action model, which amplifies the 
magnetic field in a turbulent ICM and saturates when the magnetic 
energy density reaches a certain 
fraction of the thermal energy density. This is
typically of a few percent.

Thus we assume that

\begin{equation}
  { B^2 \over 8\pi } = \eta~ {3\over 2}~ n~k_BT~~~~,
\end{equation}
where $\eta$ is gives the energy density ratio between the magnetic
field and the thermal ICM.
We also assume for simplicity that the cluster is isothermal. The
RM scatter can then be predicted from the electron density
distribution, the cluster temperature, and the parameter $\eta,$

\begin{eqnarray}
 \sigma(RM) \propto \eta^{1/2}~\Lambda^{-1}~3^{-1/2}~ T^{1/2}~ \int n_e^{1.5}~ dl \nonumber \\
 ~~~\sigma(RM) \propto \tilde{\eta}^{1/2}~ T^{1/2}~ \int n_e^{1.5}~ dl~~~~~,
\end{eqnarray}

with a factor of $3^{1/2}$ entering, because we consider the total
magnetic field energy density, but only the line of sight component affects
the RM.

By comparing a scaled $RM$ to the predicted RM,
we can obtain the unknown ratio parameter $\eta$. In Fig. 6
we plot ${RM \over  T^{1/2}}$ versus the predicted
RM assuming $\eta = 1$ for the value given on
the x-axis. The slope of this plot reflects $\tilde{\eta}^{1/2}$,
which is about $0.004 - 0.005$, shown as a
dashed line in the plot. Adopting the parameter $\Lambda \sim 10$ 
and all three spatial components of the magnetic field for
a globally isotropic configuration, this yields a value for
$\eta$ of $5 - 7.5 \times 10^{-3}$. Including additional uncertaities,
we obtain an estimate of the ratio of magnetic to thermal 
energy density of  $3 - 10 \times 10^{-3}~ (l/10kpc)^{-1/2}$.

\begin{figure}
   \includegraphics[width=\columnwidth]{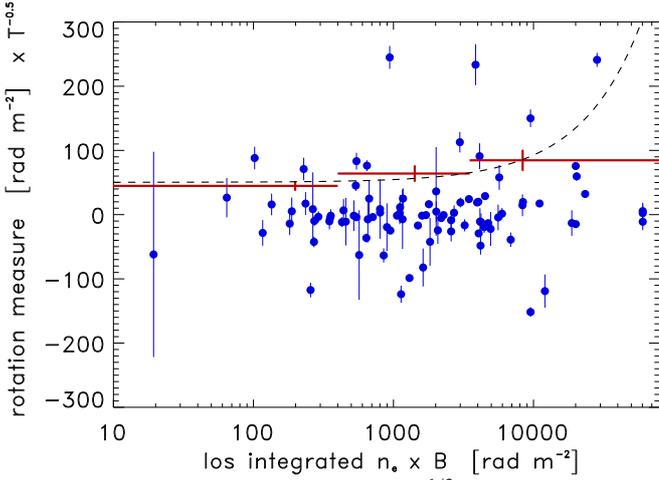}
\caption{Rotation measure scaled by $T^{1/2}$ as a function of 
the line of sight integrated electron density $\times$ magnetic 
field strength. The red bars show the
scatter of the scaled rotation measure in three bins with 
uncertainties. The dashed line shows the best fit to the scatter 
of the rotation measure. 
}\label{fig6}
\end{figure}

\section{Summary and conclusion}

Comparing Faraday RMs of polarised extragalactic 
radio sources in the line of sight of galaxy clusters with 
RM measurements made outside the projected cluster regions, we find
a clear excess of the standard deviation of the RM values 
in the cluster areas. The number of RM measurements 
is a factor of four larger than that in our
previous study reported by Clarke et al. (2001). From the values
given in Table 1, we deduce a significance of the signal 
above the background of 7 - 8$\sigma$, as determined
in the studies labelled A and C in the table.

Given these improved statistics, we can now better correlate 
RM measurements with physical parameters of the clusters. We find
that the scatter in the RM values clearly increases with cluster 
mass and electron column density in the line of sight as estimated
for our fiducial cluster model.
From the correlation of the RM scatter with electron column 
density, we deduce a typical magnetic field strength of
 $ 2 - 6~ (l/10kpc)^{-1/2} \mu$G, with the implicit assumption 
that the magnetic field is constant throughout the cluster.
This result implies that the energy density in the magnetic field
and its pressure is typically a few percent of the thermal 
energy density. In an alternative model for the ICM magnetic field
in which we assumed that the magnetic field energy density is
proportional to the thermal energy density, we
found that the energy in the magnetic field is only several 
per mille of the thermal energy. The difference in the two results
shows that the outcome of the modelling depends on the way
physical quantities are averaged over the cluster volume.
For a magnetic field that correlates with the ICM energy density or
pressure, the RM effect depends with a higher power than linearity
on the density, the central region has a larger effect, and
the overall energy required for the magnetic field is lower.
A similar result has been found in the study of Murgia et al. (2004),
where the magnetic field energy was modelled by a power spectrum.
The required magnetic field strength for the detailed model was
lower than that for the assumption of a homogeneous magnetic field.
Therefore we conclude that from averaging over the entire cluster
volume (out to $r_{500}$), the magnetic field energy is 
slightly lower than $1\%$ of the thermal energy. 

This has important implications for cluster mass measurements
based on the hydrostatic equilibrium of the ICM plasma, where
neglecting the magnetic field pressure would lead to
an underestimate of the cluster mass. For total mass estimates out to a radius of $r_{500}$, for example, the magnetic field
at larger cluster radii matters, and according to the above
discussion, we expect the magnetic field there to contribute less 
than about $1\%$ to the ICM pressure,
making it a negligible effect in the error budget of current cluster mass measurements. 

Finally, and in this context,
it is interesting to compare clusters with 
the interstellar medium of our Galaxy. While for our Galaxy
we typically find that the energy density in the magnetic field
and in cosmic rays is comparable to the thermal energy density
(Jenkins \& Tripp 2011, Draine 2011), 
where the magnetic field in the Milky Way disk  
is about 0.6 $\mu$G (e.g. Heiles \& Crutcher 2005, Kronberg 2016), 
and the magneto-ionic
thickness has been estimated between 1 and 1.8 kpc 
(Simard-Normandin \& Kronberg 1980, Sun et al., 2008, 
Gaensler et al. 2008). In contrast, galaxy clusters appear to have a lower 
ratio of $\varepsilon$(B)/$\varepsilon$ (thermal).   

What makes these
ratios different? In clusters the driving force of the
energetic processes is gravitation in the merging of subunits to
form a cluster. This is also the main energy source to generate 
turbulence and a magnetic field. \footnote{Supermassive
black holes in the central dominant galaxy can also contribute
to the generation of turbulence in clusters. For very massive
clusters this will mostly affect the central region, however. The RM
signal that we detect comes mostly from massive clusters and 
is detected globally. In this case, we expect that the influence 
of supermassive black holes is rather limited.}
In the Galaxy the energy input into the ICM and
cosmic rays comes mostly from supernovae and to some degree also
indirectly from the differential rotation of the galaxy system.
Even though the two systems seem to have many properties
in common, the main energy source is therefore very different. For clusters
the formation process and the generation of a cluster-wide magnetic
field has been simulated in detailed magneto-hydrodynamic simulations
in a cosmological model frame (e.g. Miniati 2014),
and it was concluded that the dynamo action in a turbulent ICM can generate
a magnetic field of the strength that is observed. The magnetic field
amplification saturates, however, when the magnetic field energy density
reaches about 2 - 3 percent of the thermal energy density
(Beresnyak 2012, Miniati \& Beresnyak 2015). The way in which
the magnetic field is generated during the formation of the clusters
therefore provides a natural way to set an upper limit on the possible magnetic
field energy density. The generation of the magnetic field in our
galaxy, which may be powered to a large extent by supernovae, is not
subject to this upper limit. 

\begin{acknowledgements}
H.B. and G.C. acknowledge support from the DFG Transregio Program TR33
and the Munich Excellence Cluster ''Structure and Evolution of the Universe''.  
G.C. acknowledges support from the DLR under grant no. 50 OR 1403, and P.P.K. 
thanks the Natural Sciences and Engineering Research Council of 
Canada for support under Discovery Grant No. A5713.
H.B. and G.C. thank the University of Toronto for support during the visit. 
H.B. thanks Joachim Tr\"umper for stimulating discussions.
\end{acknowledgements}


\begin{thebibliography}{}

\bibitem[B\"ohringer (2000)]{}
B\"ohringer, H., Voges, W., Huchra, J.P., et al., 2000, ApJS, 129, 435

\bibitem[B\"ohringer (2004)]{}
B\"ohringer, H., Schuecker, P., Guzzo, L., et al., 2004, A\&A, 425, 367

\bibitem[Bo\"ohringer (2007]{}
B\"ohringer, H., Schuecker, P., Pratt, G.W., et al., 2007, A\&A 469 363

\bibitem[B\"ohringer (2010)]{}
B\"ohringer, H., Werner, N., 2010, A\&ARv, 18, 127

\bibitem[B\"ohringer (2013)]{}
B\"ohringer, H., Chon, G., Collins, C.A., et al., 2013, A\&A, 555, A30

\bibitem[Carilli (2002)]{}
Carilli, C.L. \& Taylor, G.B., 2002, ARA\&A, 40, 319

\bibitem[Chon (2012a)]{}
Chon, G., \& B\"ohringer, H., 2012, A\&A, 538, 35

\bibitem[Chon (2012b)]{}
Chon, G., B\"ohringer, H., Smith, G.P., 2012, A\&A, 548, A59

\bibitem[Clarke (2001)]{}
Clarke, T.E., Kronberg, P.P., B\"ohringer, H., 2001, ApJ, 547, L111

\bibitem[Croston (2008)]{}
Croston, J.H., Pratt, G.W., B\"ohringer, H., et al., 2008, A\&A, 487, 431
\bibitem[Dickey (1990)]{}
Dickey, J.M., Lockman, F.J., 1990, ARA\&A, 28, 215

\bibitem[Draine (2011)]{}
Draine, B., 2011, Physics of the interstellar and intergalactic medium, 
Princeton Univ. Press

\bibitem[Eilek (2002)]{}
Eilek, J.A. \& Owen, F.N., 2002, ApJ, 567, 202

\bibitem[Feretti (1995)]{}
Feretti, L., Dallacasa, D., Giovannini, G., et al., 1995, A\&A, 302, 680 

\bibitem[Feretti (1999)]{}
Feretti, L., Dallacasa, D., Govoni, F., et al., 1999, A\&A, 344, 472

\bibitem[Feretti (2012)]{}
Feretti, L., Giovannini, G., Govoni, F., et al., 2012, A\&ARv, 20, 54

\bibitem[Gaensler (2008)]{}
Gaensler, B.M., Madsen, G.J., Chatterjee, S., \& Mao, S.A., 2008, Publ. Astron. Soc. of Australia, 25, 184

\bibitem[Govoni (2001)]{}
Govoni, F., Taylor, G.B., Dallacasa, D., et al., 2001, A\&A, 379, 807

\bibitem[Heiles (2005)]{}
        
Heiles, C. \& Crutcher, R., 2005, in Cosmic Magnetic Fields, 
R. Wielebinski, R. Beck (eds.), Lecture Notes in Physics, 664, 137

\bibitem[Jenkins (2011)]{}
Jenkins, E.B., Tripp, T.M., 2011, ApJ, 734, 65

\bibitem[Kim (1990)]{}
Kim, K.-T., Kronberg, P.P., Dewdney, P.E., 1990, ApJ, 355, 29 

\bibitem[Kim (1991)]{}
Kim, K.-T., Tribble, P.C., Kronberg, P.P., 1991, ApJ, 379, 80

\bibitem[Kravtsov (2012)]{}
Kravtsov, A.V. \& Borgani, S., 2012, ARA\&A, 50, 353 

\bibitem[Kronberg (2011]{}
Kronberg, P.P. \& Newton-McGee, K.J., 2011, Publ. Astron. Soc. of Australia, 28, 171

\bibitem[Kronberg (2011]{}
Kronberg, P.P., Kothes, R., Salter, C.J., Perilat, P., 2007, ApJ, 659, 267

\bibitem[Kronberg (2016]{}
Kronberg, P.P., 2016, Cosmic Magnetic Fields, Cambridge University Press,
(in press)

\bibitem[Lawler (1982)]{}
Lawler, J.M. \& Dennison, B., 1982, ApJ, 252, 81

\bibitem[Miniati (2015a)]{}
Miniati, F., \& Beresnyak, A., 2015, Nat., 523, 59

\bibitem[Miniati (2015b)]{}
Miniati, F., 2015, ApJ, 800, 60

\bibitem[Murgia (2004)]{}
Murgia, M., Govoni, F., Feretti, L., et al., 2004, A\&A, 424, 429

\bibitem[Owen (1997)]{}
Owen, F.N. \& Ledlow, M.J., 1997, ApJS, 108, 41

\bibitem[pshirkov (2011)]{}
Pshirkov, M.S., Tinyakov, P.G., Kronberg, P.P., \& Newton-McGee, K.J., 2011, 
ApJ, 738, 192

\bibitem[Simard (1980)]{}
Simard-Normandin, M. \& Kronberg, P.P., 1980, ApJ, 242, 74

\bibitem[Simard (1981)]{}
Simard-Normandin, M., Kronberg, P.P., Button, S., 1981, ApJS, 45, 97 

\bibitem[Sun (2008)]{}
Sun, X.H., Reich, W., Waelkens, A., \& Ensslin, T.E., A\&A 477, 573

\bibitem[Taylor (2009)]{}
Taylor, A.R., Stil, J.M., Sunstrum, C., 2009, ApJ, 702, 1230

\bibitem[Taylor (2001)]{}
Taylor, G.B., Govoni, F., Allen, S.W., et al., 2001, MNRAS, 326, 2

\bibitem[Tr\"umper (1993)]{}
Tr\"umper, J., 1993, Science, 260, 1769

\bibitem[Voges (1999)]{}
Voges, W., Aschenbach, B., Boller, T., et al. 1999,  A\&A, 349, 389

\bibitem[Voit (2005)]{}
Voit, M.G., 2005, RvMP, 77, 207

\bibitem[Xu (2006)]{}
Xu, Y., Kronberg, P.P., Habib, S., et al., 2006, ApJ, 637, 19

\end{thebibliography}
\end{document}